\documentclass[graybox]{svmult}

\usepackage{type1cm}        % activate if the above 3 fonts are
\usepackage{makeidx}         % allows index generation
\usepackage{graphicx}        % standard LaTeX graphics tool
\usepackage{multicol}        % used for the two-column index
\usepackage[bottom]{footmisc}% places footnotes at page bottom

\usepackage{newtxtext}       % 

\usepackage[colorlinks,citecolor=blue,urlcolor=blue,linkcolor=blue,bookmarks=true,hypertexnames=true]{hyperref} 

\makeindex             % used for the subject index
\begin{document}

\title*{Study of neutron response using time of flight technique in ISMRAN detector}
\titlerunning{Neutron response in ISMRAN using TOF technique}

\author{R.~Dey\thanks{Email: \texttt{neuphyroni@gmail.com}}, P.~K.~Netrakanti, D.~K.~Mishra, S.~P.~Behera, R.~Sehgal, V.~Jha, and L.~M.~Pant}
\authorrunning{R. Dey et al.}

\institute{
Nuclear Physics Division, Bhabha Atomic Research Centre, Trombay, Mumbai - 400085, India
}

\maketitle

\vspace*{-32mm}

\abstract{We report the measurements of the fast neutron energy response in Indian Scintillator Matrix for Reactor Anti-Neutrinos (ISMRAN) detector consisting of an array of 9$\times$10 Plastic Scintillator Bars (PSBs) at BARC, Mumbai. ISMRAN is an above ground detector setup at $\sim$13 m from the Dhruva reactor core for the detection of reactor anti-neutrinos (${\overline{\ensuremath{\nu}}}_{e}$) via the inverse beta decay (IBD) process. The dominant sources of reactor-related background in the vicinity of the detector are high energy $\gamma$-rays and fast neutrons. Therefore, a good understanding of fast neutron response in PSB is an essential pre-requisite for suppression and discrimination of the fast neutron background from IBD events. Kinetic energies of the fast neutron were determined using the Time-of-Flight (TOF) technique, which is used to get the scintillation light yield due to recoiling proton in PSB. We also measured the fast neutron capture time distribution in ISMRAN array using a novel technique involving TOF of the measured fast neutrons. The observed characteristic neutron capture time ( $\tau$ ) of 68.29 $\pm$ 9.48 $\mu$s is in good agreement with GEANT4 based MC simulation. These experimentally measured results will be useful for discriminating correlated and uncorrelated (accidental) background events from the true IBD events in reactor ON and OFF conditions inside the reactor hall.}

\vspace*{-3mm}

\section{Introduction:}
\label{sec:1}
ISMRAN detector setup is designed to measure the yield and energy spectrum of ${\overline{\ensuremath{\nu}}}_{e}$, via the inverse beta decay (IBD) process, for monitoring the reactor thermal power and fuel evolution and also search for the existence of sterile neutrino with a mass on the order of $\sim$1 eV /$\mathrm{c^{2}}$. The excess of ${\overline{\ensuremath{\nu}}}_{e}$ events in data compared to the predictions particularly at the energy range between 5 MeV to 7 MeV in the measured positron energy spectrum will also be addressed using ISMRAN array. The detector setup at Detector Integration Laboratory (DIL) in BARC, consisted of 90 PSBs, arranged in the form of a matrix in an array of 9$\times$10 in non-reactor environment. Each PSB is wrapped with Gadolinium Oxide ($\mathrm{Gd_{2}O_{3}}$) coated on aluminized mylar foils, 100 cm long with a cross-section of 10 $\times$ 10 $\mathrm{cm^2}$~\cite{pawan}. Three inch diameter, PMTs are coupled at the both ends of each PSB for signal readout of the triggered events. The data acquisition system (DAQ), CAEN VME based 16 channels and 14 bits waveform digitizer (V1730) of high sampling frequency 500 MS/s has been used for the pulse processing and event triggering from each PSB independently. The anode signals from PMTs at both ends of a PSB are required to have a time coincidence of 20 ns to be recorded as a triggered event. The timestamped data from each PSB is then further analyzed offline using energy deposition, timing and position information to build an event. In this paper, we present the measurement of  scintillation-light yield of fast neutron due to the recoiling protons in PSB, which has been determined by converting TOF spectrum to neutron kinetic energy to get the proton recoil parametrization of PSB. Fast neutron capture time distribution in ISMRAN array has also been discussed in detail. 

\vspace*{-5mm}

\section{Results and discussion:}
\subsection{Fast neutron energy response in the ISMRAN array:}
\label{subsec:2}
\begin{figure}
\begin{center}
\includegraphics[scale=.35]{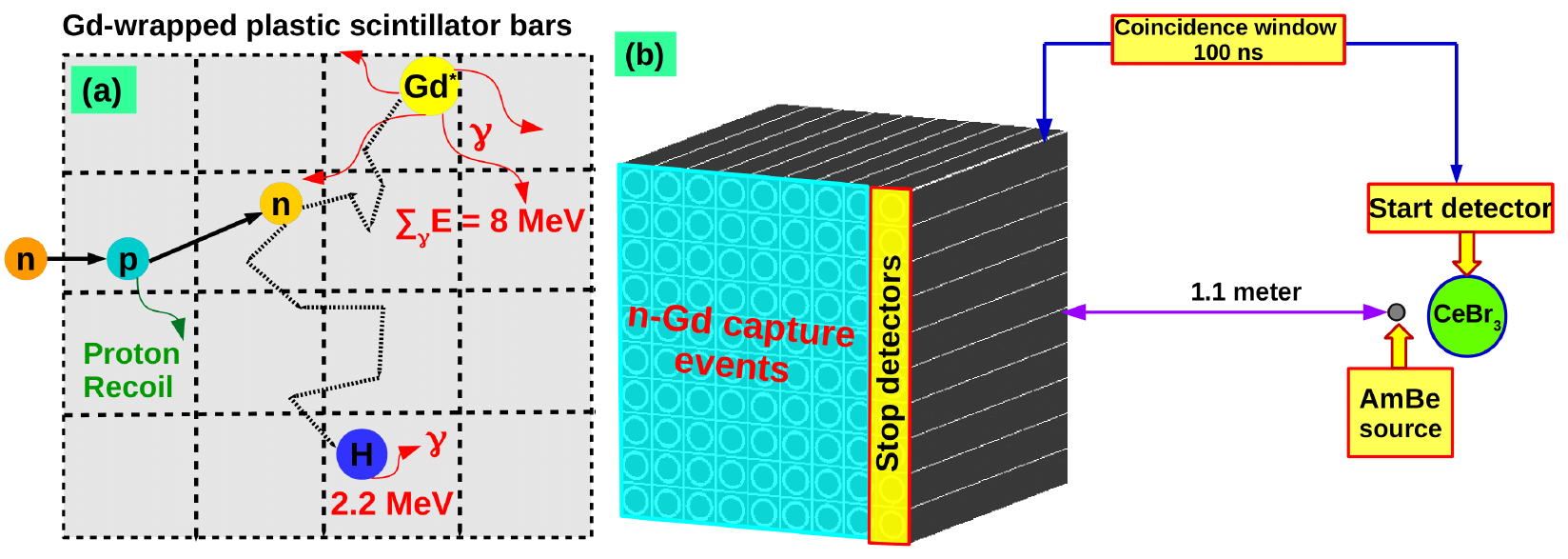}
\caption{Panel (a) shows the schematic representation of fast neutron event mimicking prompt and delayed event signatures in ISMRAN array. Panel (b) shows the schematic representation of TOF experimental set up at DIL.}
\label{fig:1}  
\end{center}   
\end{figure}

\begin{figure}[b]
\begin{center}
%\hspace{-2.9em}
\includegraphics[scale=.55]{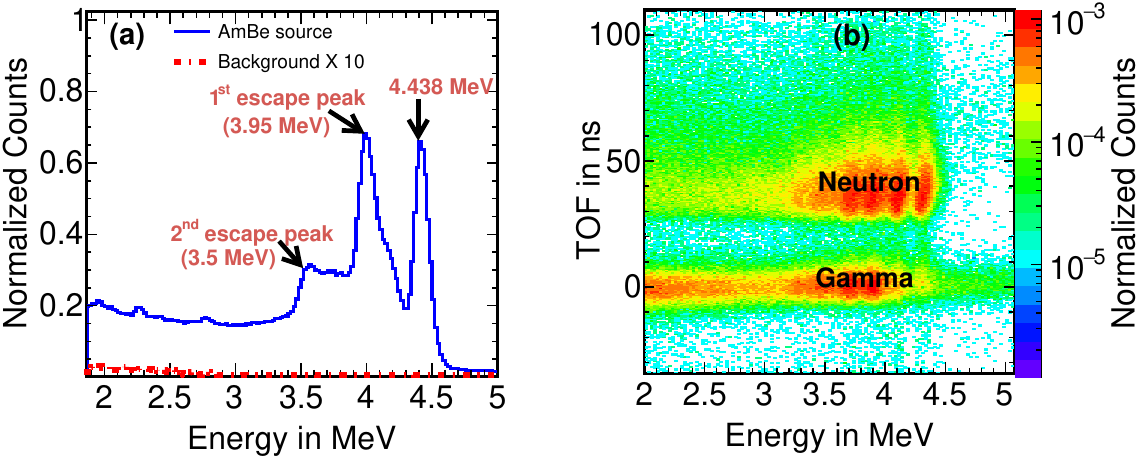}
\caption{Panel (a) shows the comparison of energy deposited untagged $\gamma$-rays spectrum from $\mathrm{{}^{241}Am}$-$\mathrm{{}^{9}Be}$ source and the $\gamma$-rays from natural background (without source) in $\mathrm{CeBr_{3}}$ detector. Panel (b) shows the TOF vs. $\gamma$ energy deposited spectrum in $\mathrm{CeBr_{3}}$ detector within the time coincidence of 100 ns between PSBs and $\mathrm{CeBr_{3}}$ detector.}
\label{fig:2} 
\end{center}    
\end{figure}

Fast neutron energy response in PSB has been measured using americium-beryllium $(\mathrm{{}^{241}Am}$-$\mathrm{{}^{9}Be})$ neutron source. Fast neutrons are produced via $\mathrm{{}^{9}Be}(\alpha, n)\mathrm{{}^{12}C}$ reaction, where the $\alpha$ particle is produced in the radioactive decays of $\mathrm{{}^{241}Am}$. In about 60$\%$ of the cases, the carbon nucleus is produced in an excited state, and emits a 4.438 MeV $\gamma$-ray in addition to the neutron. We used a 2$''$ cerium bromide ($\mathrm{CeBr_{3}}$) detector for triggering 4.438 MeV $\gamma$-ray, which provides a reference start time for the corresponding emitted neutrons. The $\mathrm{CeBr_{3}}$ detector has been placed close to the $\mathrm{{}^{241}Am}$-$\mathrm{{}^{9}Be}$ source. The ISMRAN array was located 1.1 m away from the $\mathrm{{}^{241}Am}$-$\mathrm{{}^{9}Be}$ source at the source height, as shown in Fig.~\ref{fig:1}. The $\mathrm{CeBr_{3}}$ detector is calibrated using standard radioactive $\gamma$-rays sources and the energy resolution is obtained 3.8$\%$ at 0.662 MeV. As a trigger for the start time, 4.438 MeV $\gamma$-ray has been tagged in $\mathrm{CeBr_{3}}$ detector and neutron or $\gamma$-ray as stopped time candidate is recorded at the first column (10 PSBs) of ISMRAN array. The time coincidence window between start and stop detectors is chosen to be 100 ns. By recording the start and stop time signals from $\mathrm{CeBr_{3}}$ and first column of ISMRAN array, the TOF is reconstructed for discrimination between the $\gamma$-rays and neutrons on the first column of the ISMRAN array. Figure~\ref{fig:2} (a) shows the comparison of energy deposited untagged $\gamma$-rays spectrum from $\mathrm{{}^{241}Am}$-$\mathrm{{}^{9}Be}$ source and the $\gamma$-rays from natural background (without source) in $\mathrm{CeBr_{3}}$ detector. The peaks in the $\gamma$-rays distribution in $\mathrm{CeBr_{3}}$ detector for $\mathrm{{}^{241}Am}$-$\mathrm{{}^{9}Be}$ source between 3.3 MeV to 5 MeV correspond to the neutron-associated $\gamma$-rays from the de-excitation of $\mathrm{{}^{12}C^{*}}$. As it can be seen from Fig ~\ref{fig:2}(a), a full-energy peak at 4.438 MeV corresponds to high energy $\gamma$-ray due to the de-excitation of carbon nucleus from neutron source while the corresponding first and second escape peaks appear at $\sim$3.95 MeV and $\sim$3.5 MeV, respectively. Figure ~\ref{fig:2} (b) displays the TOF distribution of fast neutron as a function of $\gamma$ energy deposition in $\mathrm{CeBr_{3}}$ detector within the time coincidence window of 100 ns. Two bands at $\sim 1$ ns and $\sim 40$ ns in the TOF distribution are due to $\gamma$-rays and fast neutron, respectively. The separation between $\gamma$-rays and neutrons is excellent and this feature is used to tag the fast neutrons on the first column of the ISMRAN array.

\begin{figure}[b]
\begin{center}
\includegraphics[scale=.55]{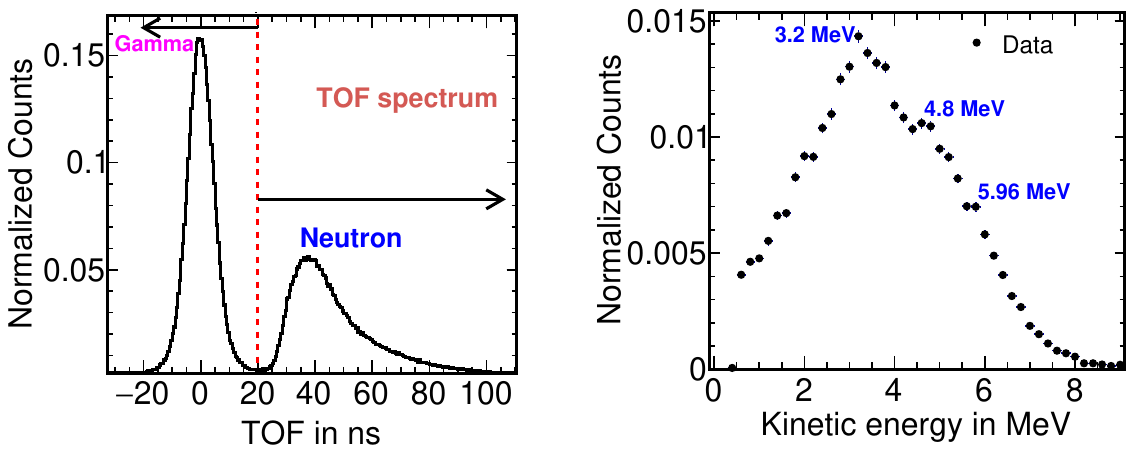}
\caption{Panel (a) shows the projected distribution of TOF spectrum for $\gamma$-ray and neutron. Panel (b) derived the kinetic energy distribution of fast neutron from $\mathrm{{}^{241}Am}$-$\mathrm{{}^{9}Be}$ source using TOF technique.}
\label{fig:3} 
\end{center}    
\end{figure}

Figure ~\ref{fig:3} (a) shows the projected distribution of TOF for gamma and neutron by tagging high energy gamma 2.0 MeV to 5.0 MeV in $\mathrm{CeBr_{3}}$ detector. A peak at $\sim 1.0$ ns correspond to the $\gamma$-rays, having width of $\sim$4.5 ns. $\mathrm{T_{0}}$, the instant of emission of the neutron from the source, was determined from the location of the $\gamma$-rays in the TOF spectrum using the speed of light (c) and the measurements of the distance (L) between the PSBs and the source. The $\gamma$-rays and fast-neutron distributions are clearly identified. The kinetic energy distribution of the neutrons can be determined by TOF spectrum using following classical expression of neutron kinetic energy. Figure \ref{fig:3} (b) shows the neutron kinetic energy distribution, derived from the TOF spectrum of tagged neutron between 20 ns to 100 ns.
\begin{equation}\label{eq:tof1}
\mathrm{ E_{n}} = \mathrm{\frac{1}{2}mV^{2}} = \mathrm{\frac{1}{2}m \left(\frac{L^{2}}{t^{2}}\right)} = \mathrm{\alpha^{2}\left(\frac{L^{2}}{t^{2}}\right)}; where,  \mathrm{\alpha = 72.3(\sqrt{ev}).\mu s/m},
\end{equation}
\begin{equation}\label{eq:tof2}
\mathrm{TOF_{n}} = \mathrm{\frac{72.3 L}{\sqrt{E_{n}}}};
\mathrm{TOF_{\gamma}} = \left(\frac{L}{c}\right).
\end{equation}

%\begin{equation}\label{eq:tof1}
%\mathrm{ E_{n}} = \mathrm{(1/2)mV^{2}} = \mathrm{(1/2)m(L^{2}/t^{2})} = \mathrm{\alpha^{2}(L^{2}/t^{2})}; where,  \mathrm{\alpha = 72.29(\sqrt{ev}).\mu s/m},
%\end{equation}
%\begin{equation}\label{eq:tof2}
%\mathrm{TOF_{n}} = \mathrm{(72.29 L)/\sqrt{E_{n}}};
%\mathrm{TOF_{\gamma}} = (L/C).
%\end{equation}

\begin{figure}[b]
\begin{center}
\includegraphics[scale=.6]{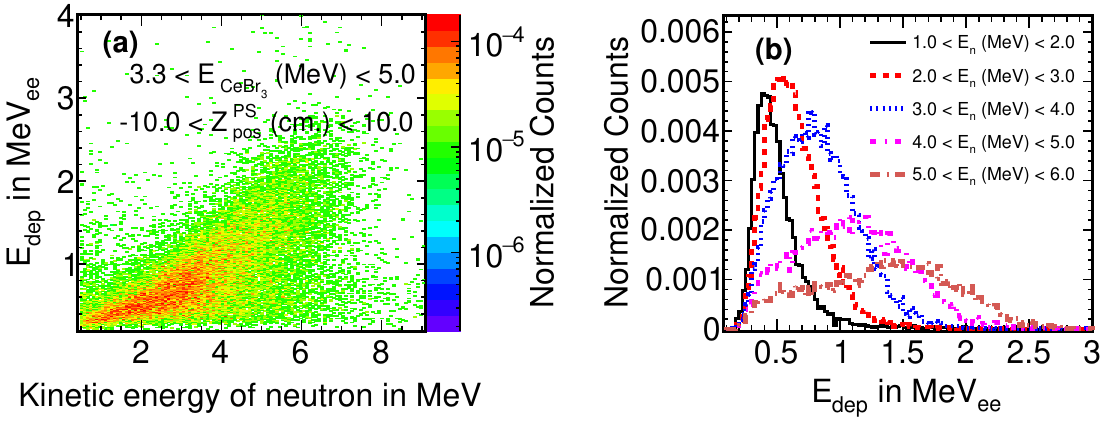}
\caption{Panel (a) shows the energy deposition ($\mathrm{E_{dep}}$) of tagged neutrons in PSBs vs. kinetic energy of tagged neutrons. Panel (b) shows the projected distribution of neutron energy deposition ($\mathrm{E_{dep}}$) for different kinetic energy bins of tagged neutrons.}
\label{fig:4}   
\end{center}  
\end{figure}

Figure ~\ref{fig:4} (a) shows the energy deposition ($\mathrm{E_{dep}}$) by fast neutrons, in $\mathrm{MeV_{ee}}$, in PSB as a function of kinetic energy of neutron derived from TOF. To minimize the contributions from the accidental natural background, $\gamma$-ray energy selection between 3.3 MeV and 5.0 MeV is made which covers the first escape, second escape and photoelectric peak for 4.438 MeV $\gamma$-ray from $\mathrm{{}^{241}Am}$-$\mathrm{{}^{9}Be}$ source. The projection of the $\mathrm{E_{dep}}$ of the neutron in PSB for different neutron kinetic energy bins is shown in Fig ~\ref{fig:4} (b). The width of the projected $\mathrm{E_{dep}}$ distribution increases with increasing kinetic energy of the neutron. This is due to the fact that exact binning in the TOF distribution for deriving kinetic energy is not possible and the smearing effect on kinetic energy is observed more prominently towards smaller TOF values which yields larger neutron kinetic energies. To reduce this smearing effect, the projection of $\mathrm{E_{dep}}$ by the fast neutrons in PSBs are plotted in bins of 1 MeV for the derived kinetic energy of the fast neutrons. Also for fast neutrons with higher energies, the multiple scattering within PSBs can be characterized by broader signal. To get the estimation of scintillation-light yield of neutron for recoiling protons in PSB, parametrization has been done between kinetic energy of neutron and $\mathrm{E_{dep}}$ due to recoiling proton in PSB with the following empirical formula, which is represented in Eqs.~\ref{eq:tof_formula}~\cite{roni},

%Figure ~\ref{fig:4} (a) shows the integrated scintillation-light yield of fast neutron in PSB as a function of kinetic energy of fast neutron. The neutron scintillation-light yield (due to recoiling protons) was determined by converting from TOF to neutron kinetic energy, binning in widths of 1.0 MeV, and filling the corresponding $\gamma$ calibrated energy. The scintillation-light yield of neutron in PSB for different neutron kinetic energy bins of width of 1.0 MeV has been shown in Fig ~\ref{fig:4} (b). The maximum energy deposited by the neutrons through recoiling proton are used to identify the structures in the measured energy from the PMTs at the ends of the PSB. Figure ~\ref{fig:5} shows the scintillation-light yield of neutron in PSB as a function of kinetic energy of neutron for different maximum energy deposited edges. To get the estimation of scintillation-light yield of neutron for recoiling protons in PSB, parameterization has been done between kinetic energy of neutron and energy deposition due to recoiling proton in PSB with the following empirical formula, which is represented in Eqs.~\ref{eq:tof_formula}~\cite{roni},

\begin{equation}\label{eq:tof_formula}
\mathrm{E_{dep}} = \mathrm{AE_{n} - B\left(1 - e^{\left(-CE_{n}^{D}\right)}\right)}
\end{equation}

%\begin{equation}\label{eq:tof_formula}
%\mathrm{E_{e}} = \mathrm{A*E_{p} - B*(1 - e^{(-C*E_{p}^{D})})}
%\end{equation}

\begin{figure}
\begin{center}
\includegraphics[scale=.35]{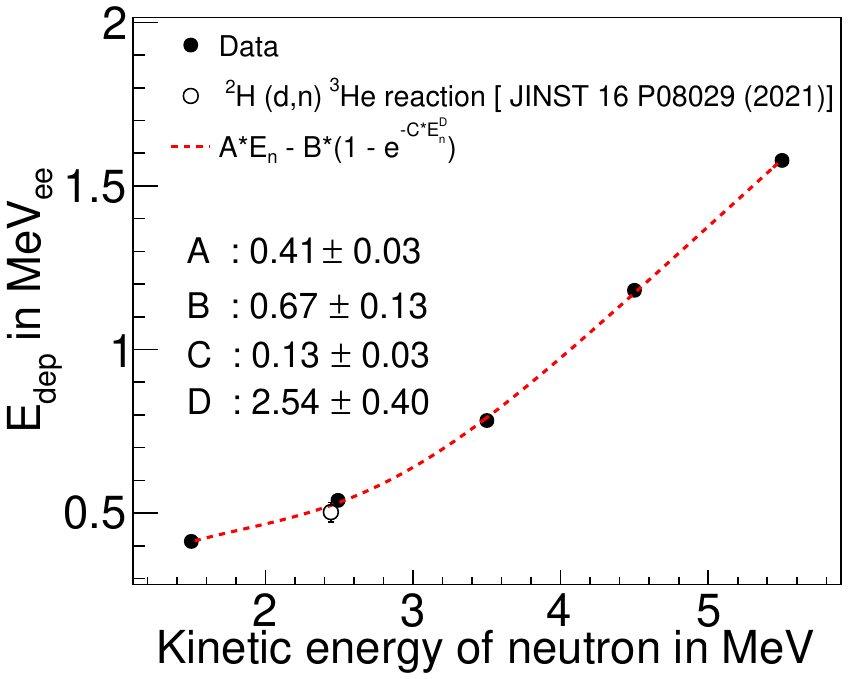}
\caption{Parametrization of $\mathrm{E_{dep}}$ due to proton recoils as a function of neutron kinetic energy in PSB.}
\label{fig:5} 
\end{center} 
\end{figure}

\vspace*{-10mm}

\subsection{Measurement of neutron capture time distribution in the ISMRAN array:}

\begin{figure}
\begin{center}
%\hspace{-1.5em}
\includegraphics[scale=.3]{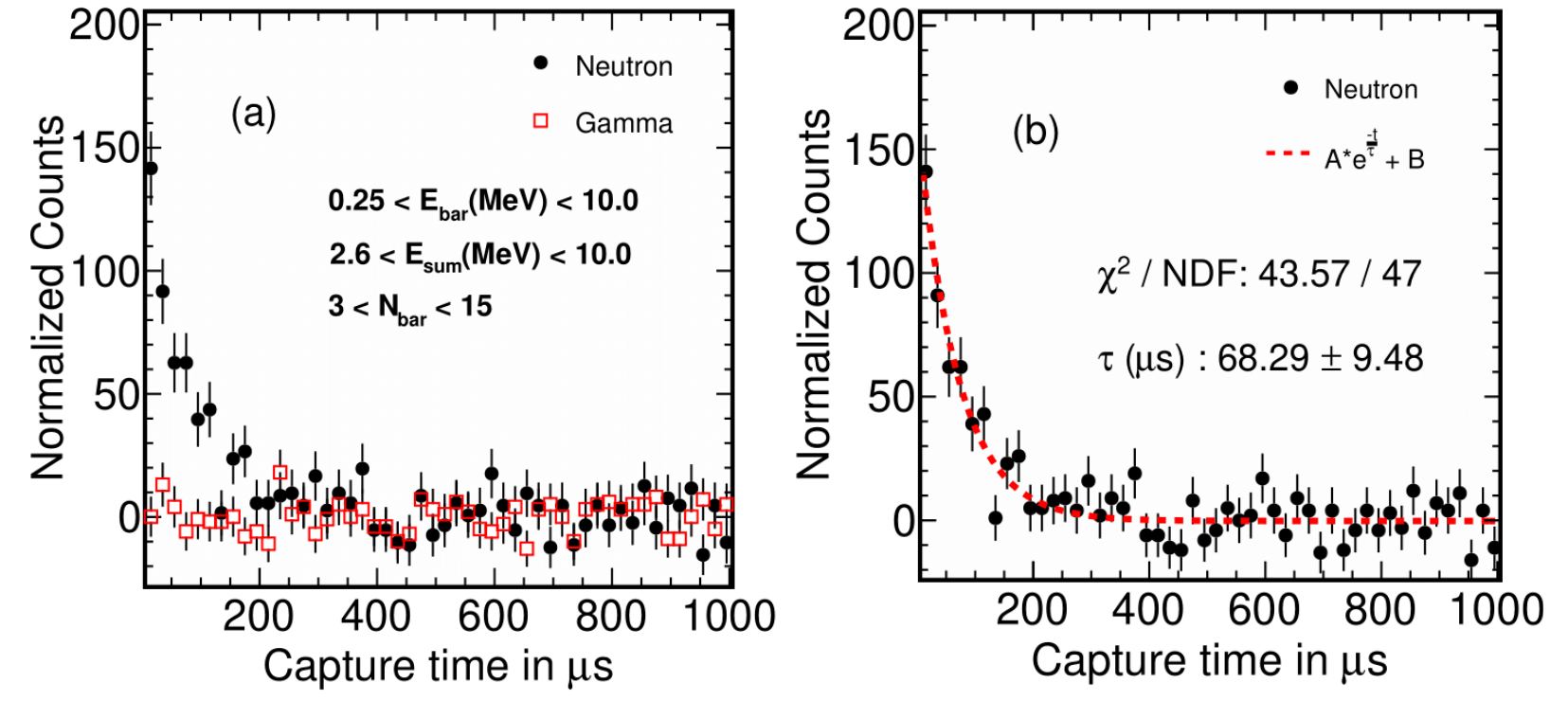}
\caption{Panel (a) shows the measured $\mathrm{\Delta T_{cap}}$ distribution for the start obtained from the first column of the ISMRAN using the tagged fast neutron or $\gamma$-ray events with stop recorded in the rest of the ISMRAN array. Panel (b) shows the fit result for $\mathrm{\Delta T_{cap}}$ distribution between tagged fast neutron from first column in ISMRAN array and stop from rest of the ISMRAN array.}
\label{fig:6}
\end{center}
\end{figure}

We have also measured the capture time distribution of fast neutron in ISMRAN detector setup. TOF technique was employed to discern between $\gamma$-rays and fast neutron from $\mathrm{{}^{241}Am}$-$\mathrm{{}^{9}Be}$ source in the first column (10 PSBs) of ISMRAN array. By tagging the neutron in this column of PSBs as a prompt event we searched the n-Gd capture event as a delayed candidate event in rest of the ISMRAN detectors within a coincidence capture time window of 1000 $\mu$s. Only those PSBs are selected for the sum energy $(\mathrm{E_{sum}})$ of delayed candidate events where the individual energy deposition in each PSB is between 0.25 MeV to 10.0 MeV, the $\mathrm{E_{sum}}$, is required to be in the energy range of 2.6 MeV to 10.0 MeV and the number of bars hit $(\mathrm{N_{bars}})$ should be in the range of 4 to 14. All the selection criteria used for the searching of stop events (delayed events) in ISMRAN array are benchmarked with GEANT4 based monte carlo simulation for fast neutron from $\mathrm{{}^{241}Am}$-$\mathrm{{}^{9}Be}$ source in ISMRAN array. Figure ~\ref{fig:6} (a) displays the n-Gd capture time ($\mathrm{\Delta T_{cap}}$) distributions of neutrons and $\gamma$ tagged events. The black solid points represents $\mathrm{\Delta T_{cap}}$ distribution for all the prompt-delayed pairs reconstructed from neutron tagged events within the time window of 1000 $\mu$s. One the other hand, the $\mathrm{\Delta T_{cap}}$ distribution for $\gamma$ tagged events (red square) shows a uniform distribution in $\Delta T_{cap}$ indicating the randomness in the prompt-delayed event pairs, which scaled with the neutron events above $\Delta T$ > 300 $\mu$s. Figure ~\ref{fig:6} (b) shows the $\mathrm{\Delta T_{cap}}$ distribution of neutron tagged events, which is fitted with a combined function consisting of an exponential term for the neutron thermalization and capture time in PSBs and a constant term representing the accidental residual background. For fast neutron, the fit results in a characteristic time ($\tau$) of 68.29 $\pm$ 9.48 $\mu$s, which is very similar to the characteristic capture time of thermal neutron for IBD delayed events. This way we have demonstrated a novel technique for the determination of the neutron capture time on Gd in ISMRAN array inspired by the data driven method.

\section{Summary:}
\label{sec:3}
The fast neutron energy response in ISMRAN detector is studied with $\mathrm{{}^{241}Am}$-$\mathrm{{}^{9}Be}$ source using TOF technique. This technique enabled the mapping of the response of the PSB to the fast neutrons as a function of their kinetic energy, which is useful to get the scintillation light yield (due to recoiling protons) parametrization for PSB. These results also indicate the capture time distribution of fast neutron are indistinguishable from those of ${\overline{\ensuremath{\nu}}}_{e}$ events. For separating the prompt IBD events from fast neutron background can be achieved by using the segmented geometry of ISMRAN array and combining energy dependent variable such as energy ratios with other topological event selection cuts in PSBs along with the implementation of an advanced machine learning algorithms.

%The capture time($\Delta T$) distribution of the triggered fast neutron events from Am/Be source in real data has been also compared with MC simulation in Geant4 after incorporating DICEBOX package for simulating the cascade $\gamma$-rays from the thermal neutron capture on Gd nucleus.

%\begin{proof}
%\smartqed
%Proof text goes here.
%\qed
%\end{proof}
%

%\begin{acknowledgement}
%If you want to include acknowledgments of assistance and the like at the end of an individual chapter please use the \verb|acknowledgement| environment -- it will automatically be rendered in line with the preferred layout.
%\end{acknowledgement}
%
%\section*{Appendix}
%\addcontentsline{toc}{section}{Appendix}

\vspace*{-3mm}

%%%%%%%%%%%%%%%%%%%%%%%% referenc.tex %%%%%%%%%%%%%%%%%%%%%%%%%%%%%%
% sample references
% %
% Use this file as a template for your own input.
%
%%%%%%%%%%%%%%%%%%%%%%%% Springer-Verlag %%%%%%%%%%%%%%%%%%%%%%%%%%
%
% BibTeX users please use
% \bibliographystyle{}
% \bibliography{}

\begin{thebibliography}{99.}%
% Use the following syntax and markup for your references if 
% the subject of your book is from the field 
% "Mathematics, Physics, Statistics, Computer Science"
%

% Journal article


%\bibitem{tof1} J. Scherzinger. et~al.: The light-yield response of a NE-213 liquid-scintillator detector measured using 2-6 MeV tagged neutrons. Nuclear Instruments and Methods in Physics Research Section A. \textbf{840}, 121-127 (2016), doi: org/10.1016/j.nima.2016.10.011.

\bibitem{pawan} P. K. Netrakanti. et~al.: Measurements using a prototype array of plastic scintillator bars for reactor based electron anti-neutrino detection. Nuclear Instruments and Methods in Physics Research Section A. \textbf{1024}, 166126 (2022), doi: org/10.1016/j.nima.2021.166126.

%\bibitem{tof} J. Scherzinger. et~al.: Tagging fast neutrons from an $\mathrm{{}^{241}Am/ {}^{9}Be}$ source. Applied Radiation and Isotopes. \textbf{98}, 74-79 (2015), doi: org/10.1016/j.apradiso.2015.01.003.

\bibitem{roni} R.~Dey. et al.: Characterization of plastic scintillator bars using fast neutrons from D-D and D-T reactions. Journal of Instrumentation. \textbf{16}, P08029 (2021), doi: https://iopscience.iop.org/article/10.1088/1748-0221/16/08/P08029.



\end{thebibliography}
%

\end{document}